\newcommand{\conlum}{$\lambda$L$_{5100 \AA}$}
\newcommand{\rblr}{R$_{BLR}$}
\newcommand{\hbeta}{H$_{\beta}$}

\documentclass[useAMS]{mn2e}
\usepackage{graphicx}

\title[The black hole mass in H~0507+164]{Mass of the black hole in
the Seyfert~1.5 galaxy H~0507+164 from reverberation mapping}
\author[C. S. Stalin et al. ]{C. S. Stalin$^{1}$\thanks{E-mail:
stalin@iiap.res.in}, S. Jeyakumar$^{2}$,
R. Coziol$^{2}$, R. S. Pawase$^{3}$,  S. S. Thakur$^{4}$ \\
$^{1}$ Indian Institute of Astrophysics, Block II, Koramangala, Bangalore 560 034, India \\
$^{2}$ Departamento de Astronomia, Universidad de Guanajuato, Guanajuato, CP36000,Mexico \\
$^{3}$ Laboratoire d'Astrophysique, Ecole Polytechnique F\'{e}d\'{e}rale 
       de Lausanne (EPFL), Observaroire de Sauverny, CH-1290, Verosix, 
       Switzerland\\
$^{4}$ Behind BSF Complex, Paloura Top, Jammu - 181124, India \\
}

\begin{document}

\date{Accepted ... Received ...; in original form ...}

\pagerange{\pageref{firstpage}--\pageref{lastpage}} \pubyear{2010}

\maketitle

\label{firstpage}

\begin{abstract}
We present the results of our optical monitoring campaign of the
X-ray source H~0507+164, a low luminosity Seyfert~1.5 galaxy at a
redshift, $z = 0.018$. Spectroscopic observations were carried out
during 22 nights in 2007, from the 21 of November to the 26 of
December.  Photometric observations in the R-band for 13 nights
were also obtained during the same period. 
The continuum and broad line fluxes of the galaxy were found to vary
during our monitoring period. The R-band differential light curve
with respect to a companion star also shows a similar variability.
Using cross correlation analysis, we estimated a
time delay of $\tau_{cen} = 3.01^{+0.42}_{-1.84}$ days (in the rest frame),
of the response of the broad H$_\beta$ line fluxes  
to the variations in the optical continuum at 5100 \AA.
Using this time delay and the width of the H$_\beta$ line,
we estimated the radius for the Broad Line Region (BLR) 
of $2.53^{+0.35}_{-1.55}  \times 10^{-3}$~parsec, and a black
hole mass of $9.62^{+0.33}_{-3.73}  \times 10^{6}$M$_{\odot}$.

\end{abstract}


\section{Introduction}

Accretion of gas onto a Super Massive Black Hole (SMBH) in the nucleus
of galaxies is believed to be the source of activity 
in Quasars and Seyfert galaxies 
(commonly known as Active Galactic Nuclei (AGNs); cf. Rees 1984).
Several studies have suggested that the mass of the SMBH
in these objects is correlated with the luminosity, mass and
velocity dispersion of the stellar spheroid of the galaxies
(Kormendy \& Richstone 1995; Magorrian et al. 1998; Ferrarese \&
Merritt 2000; Gebhardt et al. 2000; Marconi \& Hunt 2003; H\"aring
\& Rix 2004). Such correlations may imply an evolutionary
relation between the growth of the SMBH and the 
host galaxy itself (e.g. Somerville et al. 2008; Shankar et al. 2009;
Hopkins \& Hernquist 2009). 
In order to study the  dependence of the various observed phenomena of AGNs
on the black hole mass and the cosmic evolution of the black holes,
independent and reliable estimates of the mass of the black holes 
are required (e.g., Goulding et al. 2010; Rafter, Crenshaw \& Wiita 2009).

One independent method to estimate the mass of the black hole is
using the reverberation mapping technique (Blandford \& McKee 1982; Peterson 1993). 
In the optical bands, the continuum flux of some AGNs, is known to 
vary on timescales as short as hours (e.g.,
Miller, Carini \& Goodrich 1989; Stalin et al. 2004). 
If the main source of ionization of the Broad Line Region (BLR) is the continuum itself,
any variation of the continuum emission can also be seen in the broad emission lines.
However, the variations in the broad line flux will have a time lag ($\tau$) 
relative to the continuum variations, which can be interpreted 
as the light travel time across the BLR.
As a first approximation, therefore, the size of the BLR is
$R_{BLR} \le c\tau$, where $c$ is the velocity of light. 
Once the $R_{BLR}$ is obtained, the mass of the black hole can also be
estimated, using the velocity dispersion of the broad component of 
the emission lines, $\sigma_{line}$, and
assuming virial equilibrium (Peterson et al. 2004; P04); see
Peterson~2010, for a recent review).

The reverberation mapping technique has been used to make estimates
of SMBH masses over a large range of redshift. However, because the
technique is observationally taxing, as it demands an enormous
amount of telescope time, to date the BLR radius of only about three
dozen AGNs (Seyfert 1 galaxies and Quasars) have been determined
(P04; Kaspi et al. 2007; Bentz et al. 2009a; Denney
et al. 2009, 2010). Nevertheless, using these estimates a correlation
was found between $R_{BLR}$ and the optical continuum luminosity at
5100 \AA\ (Kaspi et al. 2000; Kaspi et al. 2007; 
P04; Denney et al. 2009; Bentz et al. 2009b).
The \rblr$-$\conlum\ relation can be considered well constrained
between the luminosities 10$^{43}$~erg sec$^{-1} < \lambda$L$_{5100 \AA}
< 10^{45}$~erg sec$^{-1}$. On the other hand, for luminosities below
10$^{43}$~erg sec$^{-1}$, only a handful of
sources are observed, and the estimated values of $R_{BLR}$ could
also indicate a flattening of the relation (see Fig. 2 of Kaspi et
al. 2005). This flattening would suggest a lower limit in the 
possible masses of SMBHs in galaxies. Although recent revisions of a few
sources made by Bentz et al. (2006) and Denney et al. (2009;2010) are
consistent with a continuation of the $R_{BLR}-\lambda$L$_{5100 \AA}$ relation
to lower luminosities, and consequently with no lower limit in the mass
for the SMBH, the correlation is still sparsely sampled.
Moreover, the $R_{BLR}-\lambda$L$_{5100 \AA}$ relation is very useful for estimating the
SMBH masses from single-epoch spectra  and calibrating other surrogate
relations used for black hole mass estimates 
(Vestergaard 2004; Shen et al.  2008). Therefore, 
estimates of $R_{BLR}$ for a larger number of sources are required.

The extrapolation of the known $R_{BLR}-\lambda$L$_{5100 \AA}$ relation to low
luminosities suggests that the time lag between the variations
of the broad line and that of the continuum  will be of the order of hours to
days, as compared to several months for high luminosity sources.
Thus, monitoring programs of short durations, but fast sampling,
 are required to estimate the reverberation time lags for
low luminosity sources.

In this paper, we present the optical spectroscopic and photometric
observations of a new low luminosity AGN, the X-ray source and
Seyfert~1.5 galaxy H~0507+164. Based on a reverberation mapping
campaign that lasted for about a month, during November-December
2007, we have obtained $R_{BLR}$ and estimated the mass of the SMBH.
In Section~2, the observations and data reductions are described.
The results of the analysis are given in Section~3, and the
conclusions are presented in Section~4.

\section{Observations and Reductions}

\begin{figure*}
\vspace*{-2.0cm} \hspace*{-1.0cm}
\includegraphics{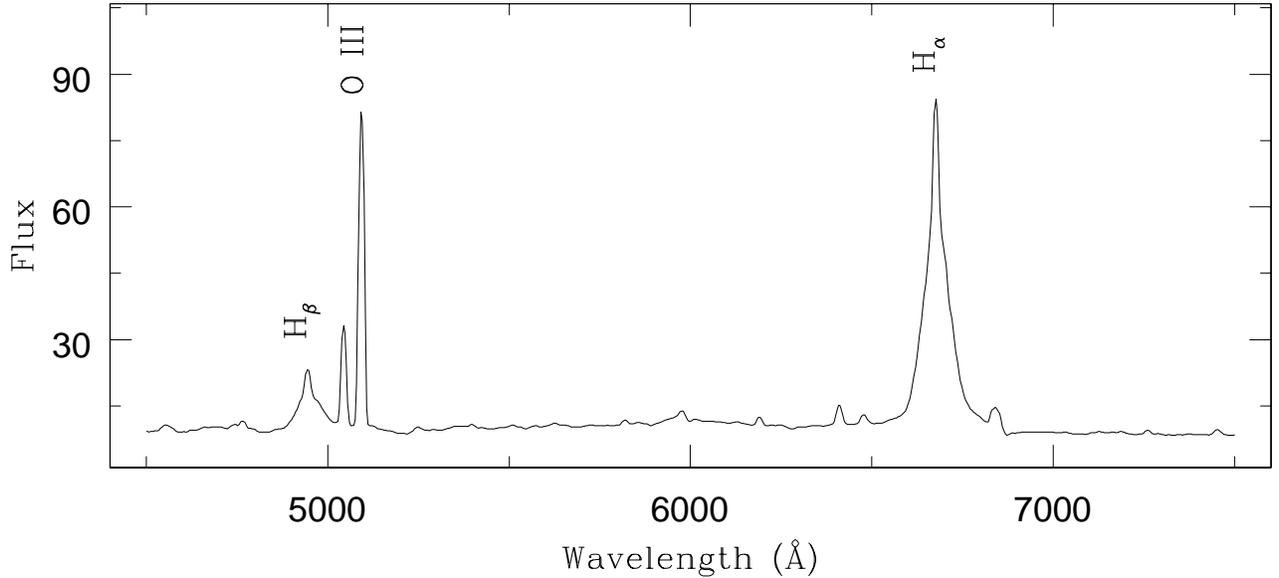}
\vspace*{-2.0cm} \caption{Average spectrum of H~0507+164 obtained by
averaging the observations of the 22 nights. The flux is in the unit of 
10$^{-16}$~erg~cm$^{-2}$~s$^{-1}$~\AA$^{-1}$.
The classification as Seyfert~1.5 is confirmed, as one can 
see both a broad and narrow component for the H$\beta$ line 
at $\lambda4861$\ \AA (cf. Osterbrock 1989).
\label{fig:meanspec}
}
\end{figure*}

Using the V\'{e}ron-Cetty \& V\'{e}ron catalogue of Quasars and Active
Galactic Nuclei (12th Ed.; V\'{e}ron-Cetty \& V\'{e}ron 2006), we have
compiled a list of nearby Seyfert~1 galaxies, which, based on the
available spectra, have a luminosity at $\lambda5100$\ \AA\
of the order of 10$^{42}$~erg sec$^{-1}$ or lower. Very few
candidates were found (mostly because of the absence of 
available spectra). The source, H 0507+164, that we selected 
for our campaign is identified in the catalogue 
of V\'{e}ron-Cetty \& V\'{e}ron as an X-ray source, with
coordinates $\alpha_{2000} = 05^h 10^m
45.5^s, \delta_{2000} = 16^d 29^m 56^s$, and is classified as a Seyfert~1.5
galaxy at a redshift of\ $z = 0.018$.

Optical spectroscopic and photometric observations of H~0507+164
were carried out in 2007 between 21 of November and 26 of
December at the 2m Himalayan Chandra Telescope (HCT), operated by
the Indian Institute of Astrophysics, Bangalore. The telescope is
equipped with a $2048 \times 4096$ CCD, coupled to the
Himalayan Faint Object Spectrograph and Camera
(HFOSC)\footnote{http://www.iiap.res.in/iao\_hfosc}. In imaging mode,
only the central $2048\times2048$ pixels region of the CCD is used.
The camera has a plate scale of $0.296$ arcsecond/pixel, which
yields a field of view of $10\times10$\ square arcmin.

\subsection{Spectroscopy}

Medium resolution spectra of the nucleus were obtained 
using a 11 arcmin $\times$ 1.92 arcsec wide slit and a grism. 
The spectra 
have a spectral range of 3800$-$6700 \AA ~with a resolution of 
$\sim$8 \AA.
The exposure time varied between 900 and 1000 seconds. The spectra
were reduced using standard procedures in IRAF\footnote{IRAF stands
for Image Reduction and Analysis Facility and is distributed by the
National Optical Astronomy Observatories, which is operated by the
Association of Universities for Research in Astronomy, Inc. under
contract to the National Science Foundation}. After bias subtraction
and flat fielding, one dimensional spectra were extracted and
calibrated, in wavelength using an FeAr lamp, and in flux using
various observations of the spectrophotometric standard star
Feige~34.  Since the observed spectra were of low S/N, for further analysis
all the spectra were smoothed to a resolution of $\sim$15 \AA. 

The standard technique of spectral flux calibration is not
sufficiently precise to study the variability of AGNs. Since even under
good photometric conditions the accuracy of spectrophotometry is not
better than 10\% (Shapovalova et al. 2008), we used a
relative calibration procedure. A first order flux calibration was first obtained in the
normal way using the standard star. Then all the spectra were
inter-calibrated relative to the spectra of one night (we choose the 21 of November),
assuming the flux of the narrow line [O~{\sc iii}]$\lambda5007$\ \AA\ is
constant. This is justified, because the Narrow Line Region (NLR) 
is much more extended (of the order of a few hundred parsecs) 
than the BLR (much less than a parsec) 
and flux variation cannot be observed in this region over short time scales 
(cf. Osterbrock 1989). Each spectra were
scaled relative to the reference spectra using the scaling algorithm
devised by van Groningen \& Wanders (1992). This algorithm uses a
chi-square, to minimize the residuals of the [O~{\sc iii}]$\lambda$5007 \AA
~line after subtraction from the reference spectrum. The mean
spectrum, averaging the 22 nights of observations, is shown in Fig.~\ref{fig:meanspec}.

\subsection{Photometry}

\begin{figure}
\hspace*{-1.5cm}
\includegraphics[width=10cm,height=7cm,angle=180]{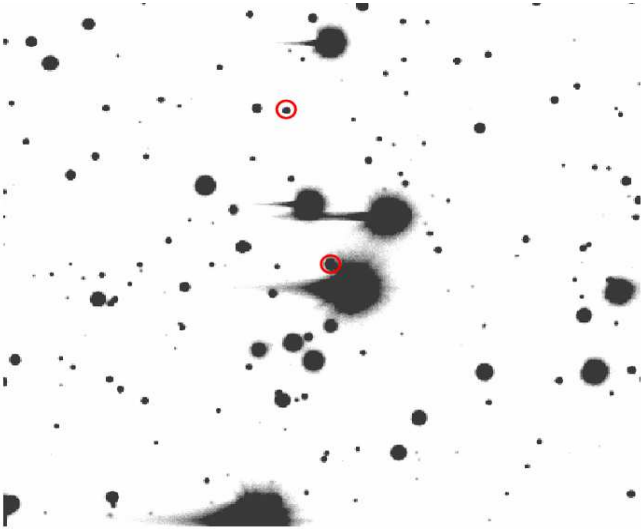}
\vspace*{0.5cm}
\caption{The R-band image of H~0507+164. The galaxy (in the center) and
comparison star (on the top) used for differential photometry are indicated
by circles. The field of view has a dimension of $10\times10$ arcmin. 
North is up and east is to the left.
Note the presence of an
extremely bright star near H~0507+164.
\label{fig:image}
}
\end{figure}

In parallel to the spectroscopic observations, R-band images were
also obtained with the HFOSC. Unfortunately, some observations
turned out to be contaminated by the light of the extremely bright
stars located near the source. As a consequence, our R-band
photometry covers only 13 of the 22 nights of the campaign.
The images were bias subtracted and flat fielded using IRAF
packages. For the rest of the reduction, calibration and analysis, 
packages in MIDAS\footnote{Munich Image Data Analysis System;
trademark of the European Southern Observatory} were used. After
removing the cosmic rays, profile fitting photometry was done using the
DAOPHOT and ALLSTAR packages. The observed R-band frame is shown in
Fig. \ref{fig:image}.

\begin{table}
 \centering
 \begin{minipage}{150mm}
\caption{Continuum and $H_{\beta}$ fluxes for the source H 0507+164
\label{table:fluxes}
}
\begin{tabular}{rccr}
\hline
Julian Date  & F$_{\lambda}\times 10^{-16}$  & $H_{\beta} \times 10^{-14}$  & $\Delta$m~~~~~~ \\
               & 5100 \AA                        & 4861 \AA                       &           \\
$-$2454420     &  (erg/s/cm$^{2}$/\AA) & (erg/s/cm$^{2}$/\AA) & (mag)~~~~~~ \\
               &                   &                   & \\
\hline
 6.434 &  6.30 $\pm$  0.07 &  5.07 $\pm$  0.05 & $-$0.022 $\pm$ 0.012 \\
 7.289 &  7.19 $\pm$  0.10 &  5.62 $\pm$  0.07 &    0.020 $\pm$ 0.015 \\
 13.429  &  7.04 $\pm$  0.11 &  5.14 $\pm$  0.07 &                      \\
 14.389 &  7.59 $\pm$  0.07 &  5.56 $\pm$  0.05 &                      \\
 15.375 &  7.98 $\pm$  0.09 &  5.11 $\pm$  0.06 & $-$0.274 $\pm$ 0.009 \\
 16.281 &  8.08 $\pm$  0.06 &  5.37 $\pm$  0.05 & $-$0.285 $\pm$ 0.008 \\
 17.376 &  8.35 $\pm$  0.08 &  5.64 $\pm$  0.05 &                      \\
 18.391 &  8.22 $\pm$  0.07 &  5.47 $\pm$  0.05 & $-$0.206 $\pm$ 0.010 \\
 19.442 &  8.51 $\pm$  0.09 &  5.52 $\pm$  0.06 & $-$0.313 $\pm$ 0.007 \\
 20.391 &  8.26 $\pm$  0.07 &  5.78 $\pm$  0.05 &                      \\
 21.421 &  8.39 $\pm$  0.09 &  5.90 $\pm$  0.07 & $-$0.364 $\pm$ 0.010 \\
 29.423 &  9.88 $\pm$  0.10 &  6.41 $\pm$  0.07 &                      \\
 30.366 &  9.90 $\pm$  0.08 &  6.38 $\pm$  0.06 & $-$0.343 $\pm$ 0.009 \\
 31.416 &  9.86 $\pm$  0.13 &  5.51 $\pm$  0.09 & $-$0.515 $\pm$ 0.006 \\
 34.183 &  9.83 $\pm$  0.09 &  6.13 $\pm$  0.06 & $-$0.422 $\pm$ 0.009 \\
 35.257 & 10.13 $\pm$  0.37 &  7.78 $\pm$  0.24 & $-$0.437 $\pm$ 0.018 \\
 36.185 & 10.30 $\pm$  0.15 &  6.62 $\pm$  0.10 &                     \\
 37.146 & 10.33 $\pm$  0.23 &  7.54 $\pm$  0.14 &                     \\
 38.092 &  8.57 $\pm$  0.42 &  8.90 $\pm$  0.26 &                     \\
 39.075 & 11.96 $\pm$  0.36 &  9.23 $\pm$  0.22 &                     \\
 40.167 & 11.53 $\pm$  0.15 &  7.64 $\pm$  0.10 & $-$0.554 $\pm$ 0.007\\
 41.326 & 10.40 $\pm$  0.13 &  6.69 $\pm$  0.08 & $-$0.536 $\pm$ 0.008\\ \hline
\end{tabular}
\end{minipage}
\end{table}

\section{Analysis}

\subsection{Lightcurves}

The lightcurves of the 
H$_\beta$ flux and the
continuum at 5100 \AA\ were obtained using the final inter-calibrated
spectra. The continuum flux in the rest-frame of the galaxy at 5100\
\AA\  was obtained using the mean flux within the observed band from 5172 to 5200\ \AA. 

\begin{figure}
\hspace{-0.2cm}
\vspace*{-1.0cm}
\includegraphics[width=8cm]{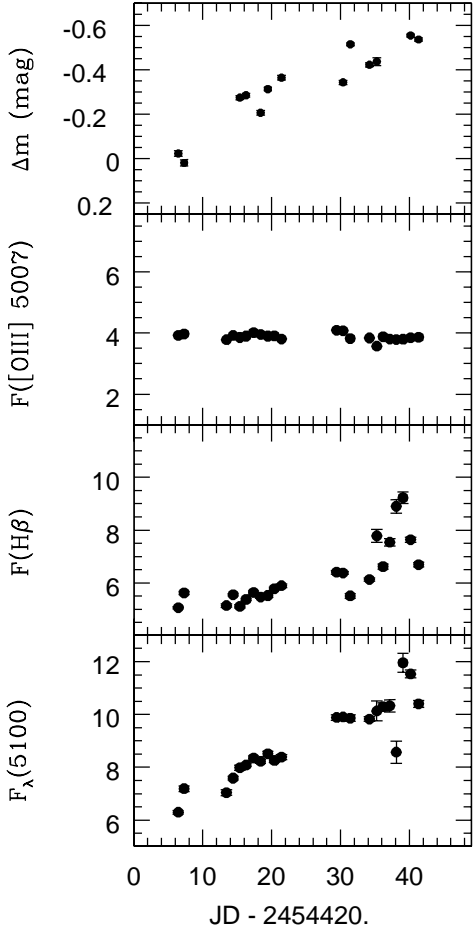}
\caption{The R-band differential lightcurve (top
panel), followed by the lightcurves for [O~{\sc iii}]$\lambda$5007 \AA,
H$_\beta$ and continuum at 5100 \AA\ are plotted. 
The fluxes are in units of
10$^{-13}$ erg s$^{-1}$ cm$^{-2}$ \AA$^{-1}$ for the
[O~{\sc iii}]$\lambda$5007 ~\AA ~line, 10$^{-14}$ erg s$^{-1}$ cm$^{-2}$
\AA$^{-1}$ for the H$\beta$ line and 10$^{-16}$ erg s$^{-1}$
cm$^{-2}$ \AA$^{-1}$ for the continuum.
\label{fig:lightcurve}
}
\end{figure}

The H$_\beta$ emission line fluxes were obtained 
by integrating the emission profile in the band spanning 4884$-$5012 \AA,
after subtracting a continuum. 
An average of the mean fluxes in the regions on the blue (4808$-$4852 \AA) and red (5012$-$5024 \AA) 
sides of the H$_\beta$ line was used as the continuum below the line. 
Although the measured line fluxes include both the narrow and broad components, 
any variation observed in the line fluxes can be attributed to the 
broad component only, since the narrow component is not expected to vary during
the period  of our observations.
The lightcurves for the continuum at 5100 \AA, for the H$_\beta$
and for the [O~{\sc iii}]$\lambda$5007 \AA\ lines are shown in Fig.~\ref{fig:lightcurve}. The
corresponding fluxes for the continuum and H$_\beta$ are listed in
Table~\ref{table:fluxes}. As expected from the inter-calibration procedure, 
the lightcurve for [O~{\sc iii}]$\lambda$5007 \AA\ is
nearly constant. On the other hand, both the continuum at 5100 \AA\
and the H$_\beta$ flux are observed to vary.

The observed R-band differential instrumental
magnitudes between the galaxy and the comparison star (marked in
Fig.~\ref{fig:lightcurve}) are also given in Table~\ref{table:fluxes}. The R-band differential lightcurve 
plotted in Fig.~\ref{fig:lightcurve} (top panel) closely follows  the lightcurves of 
the continuum at 5100 \AA\ and H$_\beta$. 
The average flux at 5100 \AA\ is $9.03\pm
1.47\times 10^{-16}$ erg cm$^{-2}$s$^{-1}$\AA$^{-1}$, which
corresponds to $\lambda$L$_{5100 \AA}$ of $3.4\pm0.55\times10^{42}$ erg
s$^{-1}$, for the cosmological parameters H$_0 = 70$ km s$^{-1}$\
Mpc$^{-1}$, $\Omega_m =0.3$ and $\Omega_\lambda =0.7$.

The variability of the light curves are characterised by the parameters,
excess variance, F$_{var}$ and the ratio between the maximum and minimum flux of the
light curves, R$_{max}$ (Rodriguez-Pascual et al. 1997; Edelson et al. 2002).
The continuum and H$_\beta$ line have F$_{var}$ of 0.16 and 0.18
and R$_{max}$ of 1.9$\pm$0.020 and 1.8$\pm$0.016 respectively.
These values are within the range of values found by other variability studies of AGNs
(cf. P04; Denney et al. 2010).

\subsection{Time Lag}

The time lag between the variations of the
continuum flux and the variations of the H$_\beta$ emission 
can be determined by cross-correlating the two light curves. 
For cross correlation analysis, the method of interpolated 
cross-correlation function (ICCF; Gaskell \& Sparke 1986; Gaskell \& Peterson 1987) 
and the method of discrete correlation function (DCF; Edelson \& Korlik 1998) 
were used. 
Although both ICCF and DCF methods produce similar results (White \& Peterson 1994),
the interpolation of the light curve during the period of gaps required by the ICCF method 
might not be a reasonable approximation to the behavior of the light curves.
Thus the DCF method is preferable for data with large gaps (Denney et al. 2009). 
For comparison, we obtain the results using both the methods.

\begin{figure}
\hspace{-0.2cm}
\includegraphics[width=8cm, height=8cm]{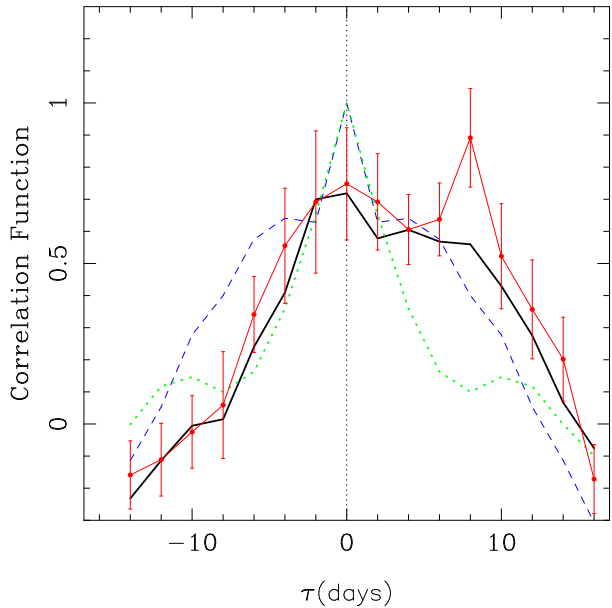}
\vspace*{-0.2cm} \caption{The cross correlation function 
of the continuum at 5100 \AA\ and the H$\beta$ lightcurves, 
as obtained using the ICCF method, is  plotted as 
as a thick solid line.  
The auto correlation functions of the continuum at 5100 \AA\ and 
the $H_{\beta}$ line are plotted as dashed and dotted 
lines respectively. The cross correlation function obtained 
using the DCF method is plotted as a thin solid line.
}
\label{fig:ccf}
\end{figure}

The results of the cross correlation analysis are shown in Fig.~\ref{fig:ccf}.
The cross correlation function (CCF) obtained using the ICCF method is plotted 
as  a thick solid line. The auto correlation functions (ACFs)
of the continuum at 5100 \AA\ and the $H_{\beta}$ line are also shown in 
Fig.~\ref{fig:ccf} as dashed and dotted lines respectively. 
For comparison the cross correlation function obtained using the DCF method 
is also plotted as a thin solid line.
As expected, the auto correlations have zero time lags. On the other hand, the time lag
in the cross correlation curve is clearly noticeable as an overall shift 
to the right. The position of the maximum in the CCF provides an
estimate of the time lag between the continuum and the H$\beta$ line.
The maximum was however determined using the centroid, which gives a better estimate
for noisy CCFs, rather than the peak, using the formula:
\begin{equation}
\tau_{cen} = \frac{\sum_{i} \tau_i CCF_i}{\sum_{i} CCF_i}
\end{equation}

The estimate of the centroid includes all the points that are within
50\% of the peak value of the CCF. 
Based on this cross correlation analysis, a statistically significant
centroid and the associated uncertainty were obtained using a 
bootstrap technique that introduces  effects of randomness in fluxes and 
sampling of the light curve (cf. P04).  A method to carry out Monte-Carlo (MC)
simulation using the combined effects of flux randomization (FR) 
and the random subset selection (RSS) procedures is described in Peterson et al. (1998). 
Additional improvements as suggested by Welsh (1999) are summarised in P04,
which we use for our analysis.

\begin{figure}
\includegraphics[width=8cm,height=8cm]{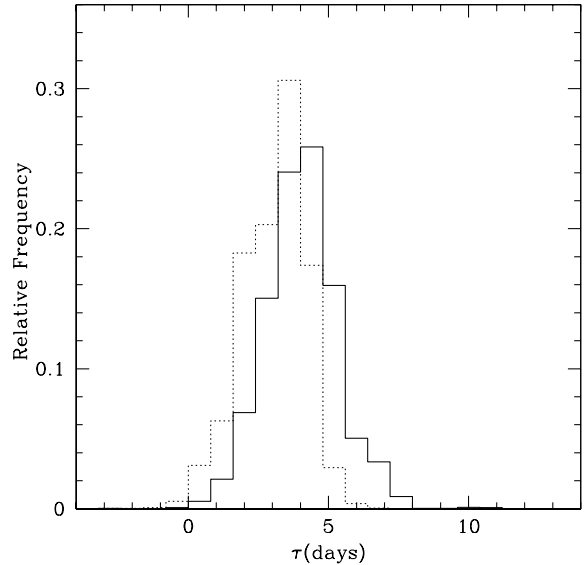}
\caption{Histogram of the cross-correlation centroids obtained using
the FR/RSS realisations. The solid and dotted lines represent the 
cross-correlation centroid distributions (CCCDs) determined using 
the ICCF and DCF respectively, for a bin size of 2 days.
\label{fig:ccfdist}
}
\end{figure}

First an RSS procedure was applied by randomly selecting 22 observations from the
light curve. The flux uncertainties of the multiply selected observations were 
weighted according to Welsh (1999). This light curve was given as input to the
FR procedure, where each measured value of the fluxes
are modified by adding the measured flux uncertainties multiplied with 
a random Gaussian value. 
The modified light curves were then cross-correlated and the centroids 
were determined as outlined above using CCF values above 50\% of the peak value. 
This procedure was repeated for $4000$ times, retaining only those
CCFs  whose maximum cross-correlation coefficient is large enough 
such that the correlation is significant with a confidence level of 95\% or larger. 
A cross-correlation centroid distribution (CCCD) 
was built using the above centroids and is shown in Fig.~\ref{fig:ccfdist}. 
The average value of CCCD was taken to be $\tau_{cent}$. 
Since the CCCD is non-Gaussian (cf. Peterson et al. 1998),
the upper and lower uncertainties in $\tau_{cent}$ were determined such 
that 15.87\% of CCCD realizations have  $\tau > \tau_{cent} + \Delta \tau_{up}$
and 15.87\% realizations have  $\tau < \tau_{cent} - \Delta \tau_{low}$.
This error in $\tau_{cent}$ corresponds to $\pm 1\sigma$ errors for a Gaussian distribution.

%
%
\begin{table}
\centering
 \begin{minipage}{100mm}
\caption{Estimates of centroid values obtained using the DCF and ICCF methods. 
\label{table:centroid} }
\begin{tabular}{lccccc}
\hline
       &  \multicolumn{5}{c}{Size of the bin (days)} \\
Method &  \cline{1-5} 
       &  2 & 2.5 & 3  & 3.5 & 4 \\
\hline
ICCF    & 3.14$^{+0.95}_{-1.08}$ 
         & 3.28$^{+0.67}_{-1.60}$ 
           & 3.27$^{+0.66}_{-1.64}$ 
             & 3.06$^{+0.43}_{-1.87}$ 
               & 3.05$^{+0.40}_{-1.77}$ \\
\\
DCF    & 3.91$^{+1.30}_{-1.07}$ 
         & 4.30$^{+1.46}_{-1.44}$ 
           & 4.19$^{+1.77}_{-0.86}$ 
             & 4.11$^{+1.19}_{-1.49}$ 
               & 4.03$^{+0.94}_{-1.40}$ \\
\hline
\end{tabular}
\end{minipage}
\end{table}

The centroid time lags obtained using different cross correlation methods 
for different bin sizes are given in Table~\ref{table:centroid}.
The variations due to  different bin sizes are within the error bars. 
The DCF method gives a mean time lag of about 4 days, 
whereas the ICCF method gives a mean value of about 3 days.  
This difference is of the order of the estimated errors on the time lags.
Considering these uncertainties, both the DCF and ICCF methods
give time delays that are consistent with each other. 
This suggests that the estimated time lag is not a spurious result
for the sampling of the light curves presented here, 
and the results of the ICCF methods are reliable. 
To be conservative, for further calculations we use the time lag 
with the largest scatter corresponding to a bin size of 3.5 days,
obtained using the ICCF method. 

Based on this analysis the average observed frame time lag
between the H$_{\beta}$ and the $\lambda$5100 \AA\ continuum light-curves
was found to be $3.06^{+0.43}_{-1.87}$ days. 
After correcting for the time dilation effects using the redshift of the source, 
we found a time lag of $3.01^{+0.42}_{-1.84}$ days in the rest frame of the source. 

The wavelength coverage of our observations, also include the H$_\alpha$ line, 
and a time lag using the H$_\alpha$ line can also be estimated.
However, by repeating the analysis procedure presented here, 
a reliable time lag using H$_\alpha$ line could not be found, 
because the correlation curves are too noisy. 
This may be due to the shorter duration of the observations  
and  the relative flux calibration 
procedure using the [O~{\sc iii}] $\lambda5007$\ \AA\ line situated 
much farther apart in wavelength from H$_\alpha$
being unreliable (Grier et al. 2008). 
Unfortunately, the nearby doublet [S~{\sc ii}] $\lambda\lambda$ 6716,6731 \AA 
~narrow lines,
which could be used for the relative calibration of H$_\alpha$, are too weak. 
Thus we do not estimate the time lag using the H$_\alpha$ line.

\begin{figure*}
\hspace*{-0.6cm}
\includegraphics[width=16cm, height=8cm]{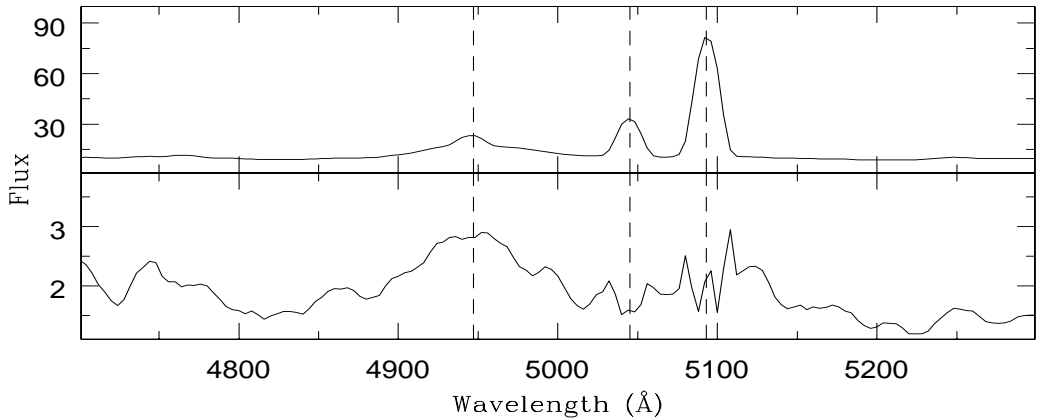}
\vspace*{-0.5cm} \caption{A zoomed up version of the mean spectrum
(top panel) and corresponding rms spectrum (bottom panel). The flux is in 
the unit of 10$^{-16}$~erg~cm$^{-2}$~s$^{-1}$~\AA$^{-1}$. The rms
spectrum shows negligible contribution from the two [O~{\sc iii}] narrow
lines.
\label{fig:rmsspec}
}
\end{figure*}

\subsection{Line width}

In order to relate the time lag to the mass of the black hole, 
an estimate of the line width, the line dispersion $\sigma_{line}$, 
of the broad emission component of H$_\beta$, is required. 
Following P04, it is relatively straight forward and more practical to 
measure $\sigma_{line}$, the second moment of the profile, 
directly from the root mean square (rms) spectrum. 
Indeed, in the rms spectrum the constant components, 
or those that vary on timescales much longer than the
duration of the observation vanish, thus largely obviating the
problem of de-blending the lines. To obtain the rms spectrum, all the observed
spectra were combined using the formula:
\begin{equation}
S(\lambda)  = \left\{  \frac{1} {{N - 1}}\sum\limits_{i = 1}^N 
{\left[
{F_i \left( \lambda  \right) - \bar F\left( \lambda  \right)}
\right]}^2   \right\}^{1 \over 2}
\end{equation}

In Fig.~\ref{fig:rmsspec}, we show the root mean square (rms) spectrum. 
It can be seen that the two narrow [O~{\sc iii}] lines have
almost completely disappeared in the rms spectrum. 

\begin{figure}
\hspace{-0.2cm}
\includegraphics[width=8cm,height=8cm]{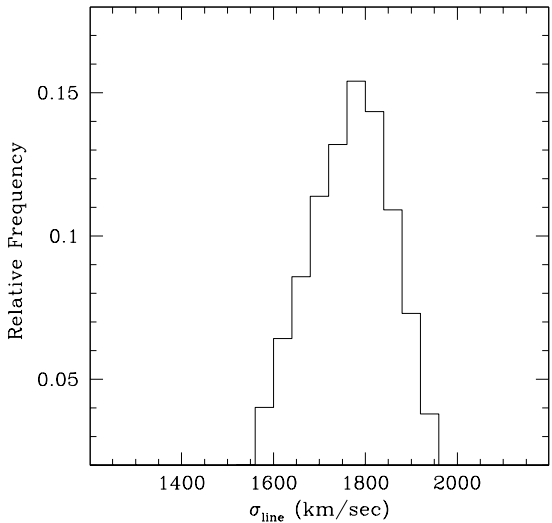}
\caption{Histogram of the estimates of the $\sigma_{line}$
using the bootstrap method described in the text.
\label{fig:sigmadist}
}
\end{figure}

The mean value of  $\sigma_{line}$ corrected for the
instrumental response of the spectrograph  and the associated uncertainty
were obtained following the bootstrap method described in P04. 
From our observed 22 observed spectra, we randomly selected 22 spectra, irrespective
of whether a particular spectrum has already been selected or not. Since
some of the spectra were selected multiple times, the mean value
of the resultant number of spectra were smaller by 8. 
These randomly selected spectra were then used to 
construct an rms spectra from which $\sigma_{line}$ was measured and corrected
for the instrumental resolution of the spectrograph.
This procedure was repeated 10000 times and the mean and standard deviation 
of these realisations  are taken as $\sigma_{line}$ and its uncertainty respectively.
A distribution of $\sigma_{line}$ values obtained from the bootstrap method 
is also shown in Fig~\ref{fig:sigmadist}. 
We thus estimate a line dispersion of $\sigma_{line} = 1725\pm105$\ km sec$^{-1}$.

\subsection{$R_{BLR}$ and the mass of the black hole}

\begin{figure}
\hspace*{-0.6cm}
\includegraphics[width=9.4cm,height=7cm]{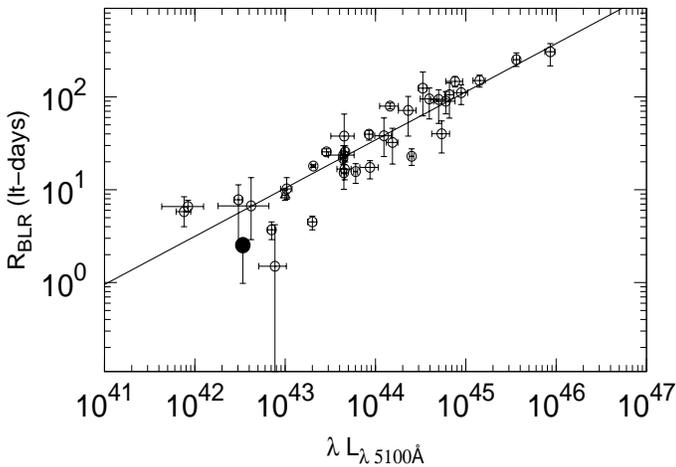}
\caption{The radius of the BLR vs the continuum luminosity
at 5100 \AA. The open circles are from Bentz et al. (2009b). 
The open triangle is the source Mrk 290 reported by Denney et al. 2010. 
The solid line (with a slope of 0.519) is the fit obtained by Bentz et al. (2009b). 
The new measurement of this work is shown as a filled circle. 
\label{fig:rlum}
}
\end{figure}

Using the rest frame time delay, the radius of the BLR is 
estimated to be $R_{BLR} = 2.53^{+0.35}_{-1.55} \times 10^{-3}$ pc. 
In Fig.~\ref{fig:rlum} we show the measurement of $R_{BLR}$ and
$\lambda$L$_{5100 \AA}$ luminosity of the source presented here
along with the most updated data set given by Bentz et al. (2009b) and 
the additional source, Mrk 290, given by Denney et al. (2010). 
The solid line is the relation obtained by Bentz et al. (2009b). 
From this figure it can be seen that our H$_{\beta}$ measurement lags are in 
agreement with the known $R_{BLR}-L$ relationship.

The mass of the black hole was estimated using the formula in P04: 
\begin{equation}
M_{BH} = f\frac{R_{BLR} {\Delta V}^2}{G} 
\end{equation}
where $\Delta V$ is the width of the line and G is the gravitational
constant. The parameter $f$ is a scaling factor, which takes into account the geometry and kinematics
of the BLR. Onken et al. (2004) found an empirical value of $f =
5.5$, using a sample of AGNs having both reverberation based black
hole masses and host galaxy bulge velocity dispersion
($\sigma_{\ast}$) estimates. This value relies on the assumption that both
AGNs and quiescent galaxies follow the same M$_{BH}$-$\sigma_{\ast}$
relationship (Ferrarese \& Merritt 2000; Gebhardt et al. 2000).
For this particular scaling, the  appropriate velocity width $\Delta V$ is
the line dispersion in the rms spectrum  $\sigma_{line}$ (Bentz et al. 2008).
Adopting the Onken et al. (2004) scaling factor and the $\sigma_{line}$ measured
from our observations,  we estimated the
mass of the SMBH in H~0507+164 to be $M_{BH} = 9.62^{+0.33}_{-3.73}  \times
10^{6}$\ M$_{\odot}$. 

\section{Conclusion}
We present for the first time monitoring observations of the X-ray
source and Sy~1.5 galaxy H~0507+164, spanning a time period of about
one month. We have obtained 22 nights of spectra during this period, 
with a mean sampling time of about 1.6 days. We measured an observed frame
time lag of about $3.06^{+0.43}_{-1.87}$ days between the changes in the
H$_\beta$ emission line flux and the changes in the
continuum flux at 5100 \AA. After correcting for the redshift, we find a
corresponding time lag of $3.01^{+0.42}_{-1.84}$ days in the rest frame
of the source. From this measured time lag we deduced a size for
the BLR of $R_{BLR} = 2.53^{+0.35}_{-1.55} \times 10^{-3}$ parsec and
estimated a black hole mass of $9.62^{+0.33}_{-3.73} \times 10^{6}$
M$_{\odot}$. Our estimate of \rblr\ using the measured lag of \hbeta\
is in agreement with the \rblr$-$\conlum\ relationship
shown by Bentz et al. (2009b).

\section*{Acknowledgments}
We thank the anonymous referee for his/her valuable comments that helped
to improve the presentation significantly.
We also thank B. M. Peterson and R. W. Pogge for kindly providing us
with the relative flux scaling program. The support provided by the staff at
the Indian Astronomical Observatory, Hanle and CREST, Hoskote is
also acknowledged.

\label{lastpage}

\begin{thebibliography}{99}
\bibitem{} Bentz M.C. et al. 2006, ApJ, 651, 775
\bibitem{} Bentz M.C. et al. 2008, ApJ, 689, L21
\bibitem{} Bentz M.C. et al. 2009a, ApJ, 705, 199
\bibitem{} Bentz M.C. et al. 2009b, ApJ, 697, 160
\bibitem{} Blandford R. D., McKee, C. F. 1982, ApJ, 255, 419
\bibitem{} Denney K.~D. et al. 2009, ApJ, 702, 1353
\bibitem{} Denney K.~D. et al. 2010, ApJ, 721, 715
\bibitem{} Edelson R.A., Krolik J.H. 1988, ApJ, 333, 646
\bibitem{} Edelson R., Turner T.~J., Pounds K., Vaughan S., Markowitz A., Marshall, H., Dobbie P., \& Warwick, R.\ 2002, ApJ, 568, 610 
\bibitem{} Ferrarese L., Merritt D. 2000, ApJ, 539, L9
\bibitem{} Gaskell C. M., Sparke L. S., 1986, ApJ, 305, 175
\bibitem{} Gaskell C. M., Peterson B.M., 1987, ApJS, 65, 1
\bibitem{} Gebhardt K. et al. 2000, ApJ, 539, L13
\bibitem{} Goulding A.~D., Alexander D.~M., Lehmer B.~D., Mullaney J.~R.\ 
2010, MNRAS, 406, 597 
\bibitem{} Grier C.~J., et al.\ 2008, ApJ, 688, 837 
\bibitem{} H\"aring N., Rix H.-W. 2004, ApJ, 604, L89
\bibitem{} Hopkins P. F., Hernquist L. 2009, ApJ, 694, 599
\bibitem{} Kaspi S., Smith P.S., Netzer H., Maoz D., Jannuzi B.T., Giveon U. 2000, ApJ, 533, 631
\bibitem{} Kaspi S., Maoz D., Netzer H., Peterson B.M., Vestergaard M., Jannuzi B.T. 2005, ApJ, 629, 61
\bibitem{} Kaspi S., Brandt W. N., Maoz D., Netzer H., Schneider D. P., Shemmer O. 2007, ApJ, 659, 997
\bibitem{} Kormendy J., Richstone D. 1995, ARA\&A, 33, 581
\bibitem{} Magorrian J. et al. 1998, AJ, 115, 2285
\bibitem{} Marconi A., Hunt L. K. 2003, ApJ, 589, L21
\bibitem{} Miller H.R., Carini M.T., Goodrich B.D. 1989, Nature, 337, 627
\bibitem{} Onken C.A., Ferrarese L., Merritt D., Peterson B.M., Pogge R.W., Vestergaard M., Wandel A. 2004, ApJ, 615, 645
\bibitem{} Osterbrock D. E 1989, in Astrophysics of Gaseous Nebulae and Active Galactic Nuclei, University Science Book
\bibitem{} Peterson B. M., Wanders I., Horne K., Collier S., Alexander T., 
           Kaspi S., Maoz D., 1998, PASP, 110, 660
\bibitem{} Peterson B. M. 1993, PASP, 105, 247
\bibitem{} Peterson B. M., et al. 2004, ApJ, 613, 682
\bibitem{} Peterson B.M., 2010, IAU Symposium, 267, 151
\bibitem{} Rafter S.~E., Crenshaw D.~M., Wiita P.~J.\ 2009, AJ, 137, 42
\bibitem{} Rees M.~J. 1984, ARA\&A, 22, 471
\bibitem{} Rodriguez-Pascual P.~M., et al.\ 1997, ApJS, 110, 9 
\bibitem{} Shankar, F., Weinberg, D. H., Miralda Escud\'e , J. 2009, ApJ, 690, 20
\bibitem{} Shapovalova A.I. et al. 2008, A\&A, 486, 99
\bibitem{} Shen Y., Greene J.E., Strauss M.A., Richards G.T., Schneider D. P. 2008, ApJ, 680, 169
\bibitem{} Somerville R. S., Hopkins P. F., Cox T. J., Robertson B. E., Hernquist L. 2008, MNRAS, 391, 481
\bibitem{} Stalin C.S., Gopal-Krishna, Sagar R., Witta P.J. 2004, JApA, 25, 1
\bibitem{} van Groningen E., Wanders I. 1992, PASP, 104, 700
\bibitem{} V\'{e}ron-Cetty M.P., V\'{e}ron P. 2006, A\&A, 455, 776
\bibitem{} Vestergaard M. 2004, ApJ, 601, 676
\bibitem{} Welsh W. F., 1999, PASP, 111, 1347
\bibitem{} White R. J., Peterson B. M., 1994, PASP, 106, 879
\end{thebibliography}
\end{document}